\documentclass[preprint,showpacs,preprintnumbers,amsmath,amssymb]{revtex4}
\usepackage{graphicx}
\usepackage{caption}
\usepackage{color}

\begin{document}

\title{STM contrast inversion of the Fe(110) surface}

\author{G\'abor M\'andi$^{a}$ and Kriszti\'an Palot\'as$^{a,b,*}$}
\affiliation{$^{a}$Budapest University of Technology and Economics, Department of Theoretical Physics,
Budafoki \'ut 8., H-1111 Budapest, Hungary\\
$^{b}$Condensed Matter Research Group of the Hungarian Academy of Sciences,
Budafoki \'ut 8., H-1111 Budapest, Hungary\\
$^{*}$Corresponding author, Email: palotas@phy.bme.hu, Tel: +3614634126}

\date{\today}

\begin{abstract}

We extend the orbital-dependent electron tunneling model implemented within the three-dimensional (3D) Wentzel-Kramers-Brillouin
(WKB) atom-superposition approach to simulate spin-polarized scanning tunneling microscopy (SP-STM) above magnetic surfaces.
The tunneling model is based on the electronic structure data of the magnetic tip and surface obtained from first principles.
Applying our method, we analyze the orbital contributions to the tunneling current, and study the nature of atomic contrast
reversals occurring on constant-current SP-STM images above the Fe(110) surface. We find an interplay of orbital-dependent
tunneling and spin-polarization effects responsible for the contrast inversion, and we discuss its dependence on the bias voltage,
on the tip-sample distance, and on the tip orbital composition.\\


\end{abstract}


\maketitle

\section{Introduction}

The scanning tunneling microscope (STM) is a great equipment to study physical and chemical phenomena on a wide range of material
surfaces in high spatial resolution. Its magnetic version, the spin-polarized STM (SP-STM) is one of the main tools for
investigating atomic scale magnetism at material/vacuum interfaces \cite{wiesendanger09review} where the inversion symmetry is
broken, giving rise to interesting magnetic phenomena. In order to explain relevant experimental findings on spin-polarized
electron transport above magnetic surfaces, advanced and efficient theoretical methods are needed \cite{hofer03rmp,hofer03pssci}.
Therefore, we propose a computationally efficient spin-polarized electron tunneling model based on the orbital-dependent
atom-superposition approach \cite{palotas12orb} to simulate SP-STM where the electronic structure of the magnetic tip and surface
is determined from first principles calculations. We apply the developed method to study the atomic contrast reversal on the
Fe(110) surface.

Even in nonmagnetic STM, at certain bias voltage and tip-sample distance ranges, constant-current STM images can show
atomic contrast inversion above flat metal surfaces, i.e., the apparent height of surface atoms can be smaller than that of the
surface hollow position, and consequently, atoms do not always appear as protrusions on the STM images. This phenomenon was widely
studied in the literature \cite{mingo96,heinze98,ondracek12,palotas12orb}. Chen theoretically explained the corrugation inversion
found on low Miller index metal surfaces by the presence of $m\ne 0$ tip states \cite{chen92}. For the particular case of the
W(110) surface, Palot\'as et al.\ \cite{palotas12orb} highlighted the role of the real-space shape of the electron orbitals
involved in the tunneling, including $m\ne 0$ tip states. On the other hand, Heinze et al.\ \cite{heinze98} pointed out that a
competition between electron states from different surface Brillouin zone parts is responsible for the corrugation inversion
effect. A similar explanation was given on the contrast inversion above the magnetic Fe(001) surface in the presence of an
external electric field \cite{heinze99}. Following the real-space electron orbital picture, we expect that beside orbital-dependent
effects \cite{palotas12orb} magnetic characteristics also play an important role in the determination of the corrugation quality
of SP-STM images on magnetic surfaces. This expectation is supported by the reported bias voltage dependent magnetic contrast
reversals in two different magnetic systems \cite{yang06,palotas11stm} obtained by using spherical tunneling models. Note that the
magnetic effect can be tuned in different ways, e.g., the magnetic contrast can be enhanced by properly adjusting the bias voltage
and/or the tip magnetization direction \cite{palotas13contrast}, or by using chemically modified magnetic STM tips
\cite{hofer08tipH}.

In the present work we report the formalism capable of studying the interplay of orbital-dependent and spin-polarized tunneling
effects within the three-dimensional (3D) Wentzel-Kramers-Brillouin (WKB) theory, and investigate the atomic contrast reversal on
the Fe(110) surface.

\section{Spin-polarized orbital-dependent tunneling model within the 3D WKB theory}
\label{sec_theory}

In SP-STM the total tunneling current can be written as the sum of a non-spin-polarized, $I_{TOPO}$, and a spin-polarized part,
$I_{MAGN}$, i.e., $I_{TOT}=I_{TOPO}+I_{MAGN}$ \cite{wortmann01,hofer03,smith04,heinze06,palotas11stm}.
We combine this concept with the orbital-dependent electron tunneling model \cite{palotas12orb} within the 3D WKB framework
based on previous atom-superposition theories \cite{tersoff85,yang02,smith04,heinze06}.
Using the nonmagnetic version of this orbital-dependent method provided comparable STM images of the W(110) surface with those
obtained by standard Tersoff-Hamann \cite{tersoff83,tersoff85} and Bardeen \cite{bardeen61} tunneling models implemented in the
BSKAN code \cite{hofer03pssci,palotas05}. The advantages, particularly computational efficiency, limitations, and the potential
of the 3D WKB method were discussed in Ref.\ \cite{palotas13fop}.

The model assumes that electron tunneling occurs through one tip apex atom only, and the transitions between this apex atom and a
suitable number of sample surface atoms are superimposed \cite{palotas11sts,palotas12orb}. Since each transition is treated within
the one-dimensional (1D) WKB approximation, and the 3D geometry of the tunnel junction is considered, the method is, in effect,
a 3D WKB approach. The electronic structure of the magnetic tip and surface is included via the atom-projected electron
density of states (PDOS) obtained by {\it{ab initio}} electronic structure calculations.
Both the charge and magnetization PDOS are necessary to describe spin-polarized tunneling \cite{palotas11stm}, and the
orbital-decomposition of the PDOS is essential for the description of the orbital-dependent tunneling \cite{palotas12orb}.

Assuming elastic tunneling and $T=0$ K temperature, the non-spin-polarized (TOPO) and spin-polarized (MAGN) parts of the
tunneling current measured at bias voltage $V$ at the $\mathbf{R}_{TIP}$ tip position can be written as
\begin{eqnarray}
\label{Eq_curtopo}
I_{TOPO}(\mathbf{R}_{TIP},V)=\int_0^V\frac{dI_{TOPO}}{dU}(\mathbf{R}_{TIP},U,V)dU,\\
I_{MAGN}(\mathbf{R}_{TIP},V)=\int_0^V\frac{dI_{MAGN}}{dU}(\mathbf{R}_{TIP},U,V)dU.
\label{Eq_curmagn}
\end{eqnarray}
The corresponding integrand can be written as a superposition of individual atomic contributions from the sample surface
(sum over $a$):
\begin{eqnarray}
\label{Eq_dIdUtopo}
&&\frac{dI_{TOPO}}{dU}(\mathbf{R}_{TIP},U,V)\\
&=&\varepsilon^{2}\frac{e^{2}}{h}\sum_a\sum_{\beta,\gamma}T_{\beta\gamma}(E_F^S+eU,V,\mathbf{d}_{a})n_{S\beta}^{a}(E_F^S+eU)n_{T\gamma}(E_F^T+eU-eV).\nonumber\\
\label{Eq_dIdUmagn}
&&\frac{dI_{MAGN}}{dU}(\mathbf{R}_{TIP},U,V)\\
&=&\varepsilon^{2}\frac{e^{2}}{h}\sum_a\sum_{\beta,\gamma}T_{\beta\gamma}(E_F^S+eU,V,\mathbf{d}_{a})\mathbf{m}_{S\beta}^{a}(E_F^S+eU)\mathbf{m}_{T\gamma}(E_F^T+eU-eV).\nonumber
\end{eqnarray}
Here, $e$ is the elementary charge, $h$ is the Planck constant, and $E_F^S$ and $E_F^T$ are the Fermi energies of the
sample surface and the tip, respectively. The $\varepsilon^{2}e^{2}/h$ factor gives the correct dimension (A/V) of the formal
conductance-like $dI/dU$ quantities \cite{palotas12sts}. The value of $\varepsilon$ has to be determined by comparing the
simulation results with experiments, or with calculations using more sophisticated methods, e.g., the Bardeen approach
\cite{bardeen61}. We chose $\varepsilon=1$ eV \cite{palotas12orb} that gives comparable current values with those obtained from
the Bardeen method. Note that the choice of $\varepsilon$ has no qualitative influence on the reported results.
$n_{S\beta}^{a}(E)$ and $n_{T\gamma}(E)$ are the orbital-decomposed charge PDOS functions of the $a$th
sample surface atom and the tip apex atom with orbital symmetry $\beta$ and $\gamma$, respectively. Similarly,
$\mathbf{m}_{S\beta}^{a}(E)$ and $\mathbf{m}_{T\gamma}(E)$ are the corresponding energy-dependent
magnetization PDOS vectors. These quantities are needed in Eqs.(\ref{Eq_dIdUtopo}) and (\ref{Eq_dIdUmagn}), respectively,
and they can be obtained from a suitable electronic structure calculation. In collinear magnetic systems
$n=PDOS^{\uparrow}+PDOS^{\downarrow}$ and $\mathbf{m}=m\mathbf{e}$ with $m=PDOS^{\uparrow}-PDOS^{\downarrow}$,
where $\mathbf{e}$ is the unit vector of the spin quantization axis \cite{palotas11stm}.
The corresponding total charge or magnetization PDOS is the sum of the orbital-decomposed contributions:
$n_{S}^{a}(E)=\sum_{\beta}n_{S\beta}^{a}(E)$,
$n_{T}(E)=\sum_{\gamma}n_{T\gamma}(E)$,
$\mathbf{m}_{S}^{a}(E)=\sum_{\beta}\mathbf{m}_{S\beta}^{a}(E)$,
$\mathbf{m}_{T}(E)=\sum_{\gamma}\mathbf{m}_{T\gamma}(E)$.
Note that a similar decomposition of the Green functions was employed within the linear combination of atomic orbitals (LCAO)
framework in Ref.\ \cite{mingo96}.

The sum over $\beta$ and $\gamma$ in Eqs.(\ref{Eq_dIdUtopo}) and (\ref{Eq_dIdUmagn}) denotes the superposition of the effect of
atomic orbitals of the sample surface and the tip, respectively, via an orbital-dependent tunneling transmission function:
$T_{\beta\gamma}(E,V,\mathbf{d}_{a})$ gives the probability of the electron tunneling from the $\beta$ orbital of the
$a$th surface atom to the $\gamma$ orbital of the tip apex atom, or vice versa. The transmission probability depends on the
energy of the electron $E$, the bias voltage $V$, and the relative position of the tip apex and the $a$th sample surface atom
$\mathbf{d}_{a}=\mathbf{R}_{TIP}(x,y,z)-\mathbf{R}_{a}(x_{a},y_{a},z_{a})$. In our model we consider
$\beta,\gamma\in\{s,p_y,p_z,p_x,d_{xy},d_{yz},d_{3z^2-r^2},d_{xz},d_{x^2-y^2}\}$ atomic orbitals,
and the following form for the transmission function:
\begin{equation}
\label{Eq_Transmission_decomp}
T_{\beta\gamma}(E_F^S+eU,V,\mathbf{d}_{a})=\exp\{-2\kappa(U,V)d_{a}\}\chi_{\beta}^2(\vartheta_{a},\varphi_{a})\chi_{\gamma}^2(\pi-\vartheta_{a},\pi+\varphi_{a}).
\end{equation}
Here, the exponential factor corresponds to an orbital-independent transmission where all electron states are considered as
exponentially decaying spherical states \cite{tersoff83,tersoff85,heinze06}, and it depends on the distance
$d_{a}=|\mathbf{d}_{a}|$ and on the vacuum decay $\kappa(U,V)=\sqrt{2m[(\phi_{S}+\phi_{T}+eV)/2-eU]}/\hbar$. We assume an
effective rectangular potential barrier in the vacuum between the sample and the tip. $\phi_{S}$ and $\phi_{T}$ are the electron
work functions of the sample surface and the tip, respectively, $m$ is the electron's mass, and $\hbar$ is the reduced Planck
constant.
The remaining factors of Eq.(\ref{Eq_Transmission_decomp}) are responsible for the orbital-dependence of the transmission
function. They modify the exponentially decaying part according to the real-space shape of the electron orbitals involved in the
tunneling, i.e., the angular dependence of the electron densities of the atomic orbitals of the surface and the tip is considered
by the square of the real spherical harmonics $\chi_{\beta,\gamma}(\vartheta,\varphi)$. $d_{a}$, $\vartheta_{a}$, and
$\varphi_{a}$ can be calculated for each surface atom from the actual tip-sample geometry:
$d_{a}=\sqrt{(x-x_{a})^{2}+(y-y_{a})^{2}+(z-z_{a})^{2}}$, $\vartheta_{a}=\arccos([z-z_{a}]/d_{a})$,
$\varphi_{a}=\arccos([x-x_{a}]/[d_{a}\sin\vartheta_{a}])$. For more details, please see Ref.\ \cite{palotas12orb}.

From Eqs.(\ref{Eq_curtopo})-(\ref{Eq_dIdUmagn}) it is clear that the total tunneling current can be decomposed according to
orbital symmetries: $I_{TOT}(\mathbf{R}_{TIP},V)=\sum_{\beta,\gamma}I_{TOT}^{\beta\gamma}(\mathbf{R}_{TIP},V)$.
The absolute current contribution of the $\beta\leftrightarrow\gamma$ transition can be calculated as
\begin{eqnarray}
&&I_{TOT}^{\beta\gamma}(\mathbf{R}_{TIP},V)=\varepsilon^{2}\frac{e^{2}}{h}\sum_a\int_0^V T_{\beta\gamma}(E_{F}^{S}+eU,V,\mathbf{d}_{a})\\
&\times&\left[n_{S\beta}^{a}(E_{F}^{S}+eU)n_{T\gamma}(E_{F}^{T}+eU-eV)+\mathbf{m}_{S\beta}^{a}(E_{F}^{S}+eU)\mathbf{m}_{T\gamma}(E_{F}^{T}+eU-eV)\right]dU,\nonumber
\end{eqnarray}
while the relative contribution can be obtained as
\begin{equation}
\label{Eq_i_bg}
\tilde{I}_{TOT}^{\beta\gamma}(\mathbf{R}_{TIP},V)=\frac{I_{TOT}^{\beta\gamma}(\mathbf{R}_{TIP},V)}{\sum_{\beta,\gamma}I_{TOT}^{\beta\gamma}(\mathbf{R}_{TIP},V)}.
\end{equation}

Note that the method can also be combined with arbitrary tip orientations \cite{mandi13tiprot} to study tip rotation
effects on SP-STM images above magnetic surfaces, and an orbital-dependent atom-superposition model for the
spin-polarized scanning tunneling spectroscopy (SP-STS) \cite{palotas11sts,palotas12sts} can also be developed.

\newpage

\section{Computational details}
\label{sec_comp}

Using the presented spin-polarized orbital-dependent tunneling model, we study the atomic contrast reversal on the SP-STM images
of the Fe(110) surface. We particularly focus on tip effects, and consider ideal magnetic tip models with different orbital
symmetries, and a more realistic iron tip. The effect of three different tip magnetization orientations is also investigated.
We choose 27 Fe(110) surface atoms involved in the atomic superposition in combination with the ideal tips,
and 112 Fe(110) surface atoms combined with the iron tip including the simulation of SP-STM images.

We performed geometry relaxation and electronic structure calculations within the generalized gradient approximation (GGA)
of the density functional theory (DFT) implemented in the Vienna Ab-initio Simulation Package (VASP)
\cite{VASP2,VASP3,hafner08}. A plane wave basis set for electronic wave function expansion in combination with the
projector augmented wave (PAW) method \cite{kresse99} was applied, and
the exchange-correlation functional is parametrized by Perdew and Wang (PW91) \cite{pw91}.
The electronic structures of the sample surface and the tip were calculated separately.

\begin{figure*}[h!]
\includegraphics[width=0.4\textwidth,angle=0]{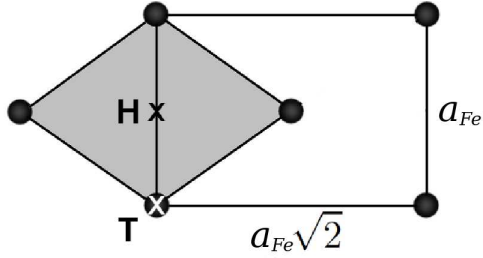}
\caption{\label{Fig1} The surface unit cell of Fe(110) (shaded area) and the rectangular scan area for the tunneling current
simulations. Circles denote the Fe atoms. The top (T) and hollow (H) positions are explicitly shown.
}
\end{figure*}

We model the Fe(110) surface by a slab of nine atomic layers with the
theoretically determined lattice constant of $a_{Fe}=2.829$ \AA, obtained at the total energy minimum of a bulk
body-centered-cubic Fe cell, in agreement with Ref.\ \cite{jiang03}. In the surface slab calculations we set up a separating
vacuum region of 24 \AA\;width in the surface normal ($z$) direction between neighboring supercell slabs to minimize slab-slab
interaction. After geometry relaxation the Fe-Fe interlayer distance between the two topmost layers is reduced by 0.39\%,
and the underneath Fe-Fe interlayer distance is increased by 0.36\% in comparison to bulk Fe.
These are in excellent agreement with the findings of Ref.\ \cite{jiang03}: -0.36\%, and +0.46\%, respectively. The size of the
in-plane magnetic moment of the surface Fe atoms is 2.50 $\mu_B$, in agreement with Refs.\ \cite{heinze98,jiang03}.
The unit cell of the Fe(110) surface (shaded area) and the rectangular scan area for the tunneling current simulations are shown
in Figure \ref{Fig1} where the surface top (T) and hollow (H) positions are explicitly indicated.
The average electron work function above the Fe(110) surface is $\phi_S=4.84$ eV
calculated from the local electrostatic potential.
We used a $41\times 41\times 5$ Monkhorst-Pack \cite{monkhorst} k-point grid for obtaining the orbital-decomposed projected
charge and magnetization electron DOS onto the surface Fe atom, $n_{S\beta}(E)$ and $\mathbf{m}_{S\beta}(E)$, respectively.

Within the presented spin-polarized orbital-dependent tunneling approach ideal magnetic tip models with particular orbital
symmetries can be considered, i.e., $\gamma_0$ orbital symmetry corresponds to the energy-independent choice of
$n_{T\gamma_0}=1(eV)^{-1}$ and $n_{T(\gamma\ne\gamma_0)}=0$. Another characteristic for ideal magnetic tips is that their absolute
spin polarization is maximal ($|P_T|=|m_T|/n_T=1$) in the full energy range, i.e., combined with a particular $\gamma_0$ orbital
symmetry: $\mathbf{m}_{T\gamma_0}=1(eV)^{-1}\mathbf{e}_T$ ($\mathbf{e}_T$ is the unit vector of the tip spin quantization axis
that is parallel to the assumed tip magnetization direction, an input parameter in our method)
and $\mathbf{m}_{T(\gamma\ne\gamma_0)}=\mathbf{0}$. We took $\gamma_0\in\{s,p_z,d_{3z^2-r^2}\}$ tip orbital symmetries.
For the ideal tips we used $\kappa(U)=\sqrt{2m(\phi_{S}-eU)}/\hbar$ vacuum decay in Eq.(\ref{Eq_Transmission_decomp}).

More realistic tips can be employed by explicitly calculating the orbital decomposition of the tip apex PDOS
in model tip geometries. In the present work a blunt iron tip is used, where a single Fe apex atom is placed on the hollow
position of the Fe(001) surface,
similarly as in Ref.\ \cite{ferriani10tip}. We took an Fe(001) slab consisting of nine atomic layers with the theoretically
determined lattice constant of $a_{Fe}=2.829$ \AA, and placed an adatom on each side of a $3\times 3$ surface cell. A separating
vacuum region of 20 \AA\;width in the surface normal direction was chosen to minimize slab-slab interaction. After geometry
relaxation of the adatom and the first surface layer performed by the VASP code, the orbital-decomposed electronic structure data
projected to the tip apex atom, $n_{T\gamma}(E)$ and $\mathbf{m}_{T\gamma}(E)$, were calculated using a $13\times 13\times 3$
Monkhorst-Pack k-point grid. We obtained the local electron work function of $\phi_T=3.96$ eV above the iron tip apex atom that
was used in the tunneling simulations.

\section{Results and discussion}
\label{sec_res}

\begin{figure*}[h!]
\includegraphics[width=1.00\textwidth,angle=0]{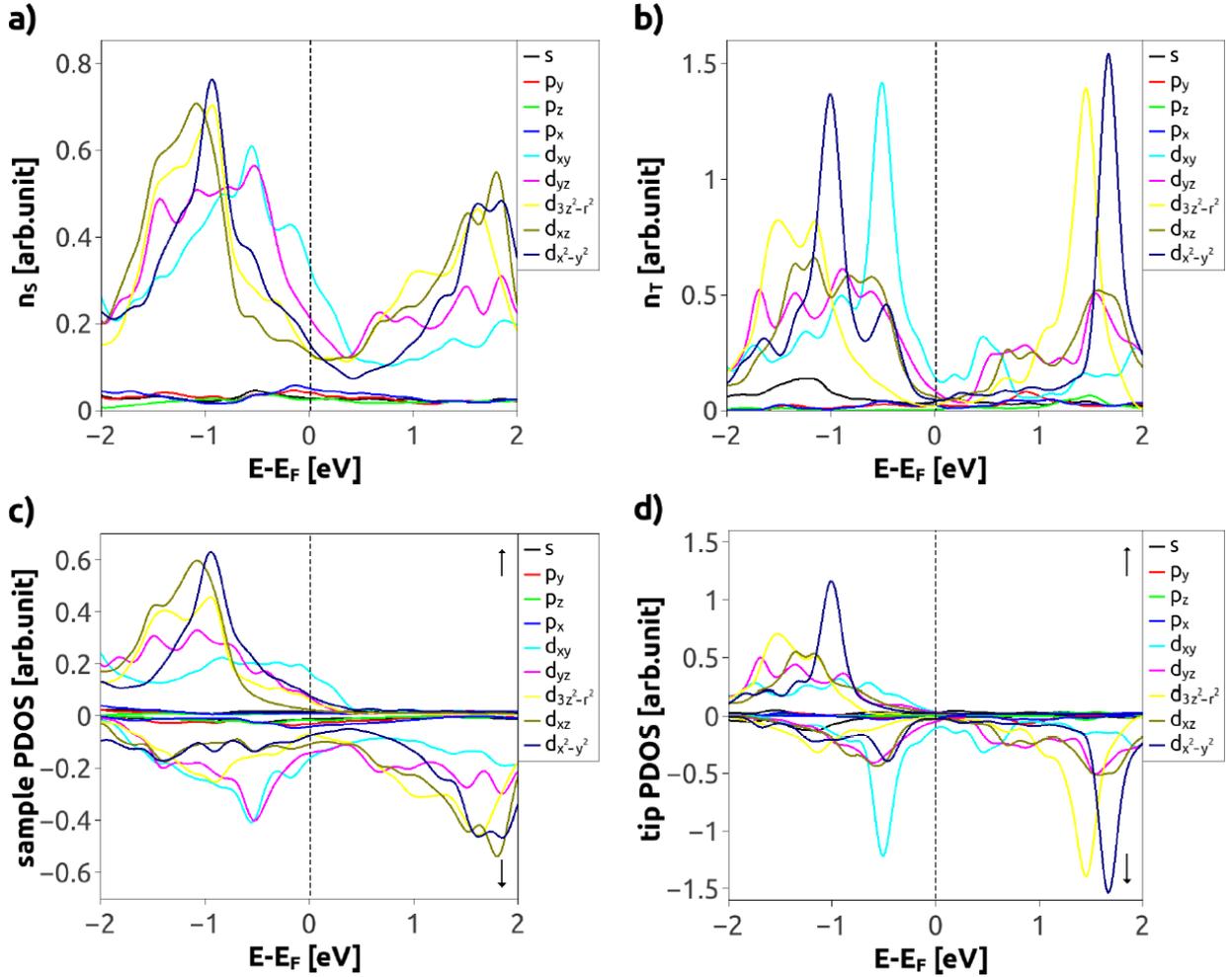}
\caption{\label{Fig2} (Color online) Orbital-decomposed projected electron density of states (PDOS) of
the Fe(110) surface atom and the iron tip apex atom.
a) surface charge PDOS: $n_{S\beta}(E)$;
b) tip charge PDOS: $n_{T\gamma}(E)$;
c) surface spin-resolved $PDOS_{S\beta}^{\uparrow,\downarrow}(E)$;
d) tip spin-resolved $PDOS_{T\gamma}^{\uparrow,\downarrow}(E)$.
Orbitals $\beta,\gamma\in\{s,p_y,p_z,p_x,d_{xy},d_{yz},d_{3z^2-r^2},d_{xz},d_{x^2-y^2}\}$ are indicated.
}
\end{figure*}

Figure \ref{Fig2} shows the energy-dependent orbital-decomposed charge ($n$) and spin-resolved
($\uparrow,\downarrow$) PDOS functions of the Fe(110) surface atom (S) and the Fe(001) tip apex atom (T) as follows:
Fig.\ \ref{Fig2}a) $n_{S\beta}(E)$, Fig.\ \ref{Fig2}b) $n_{T\gamma}(E)$, 
Fig.\ \ref{Fig2}c) $PDOS_{S\beta}^{\uparrow,\downarrow}(E)$, Fig.\ \ref{Fig2}d) $PDOS_{T\gamma}^{\uparrow,\downarrow}(E)$, with
$\beta,\gamma\in\{s,p_y,p_z,p_x,d_{xy},d_{yz},d_{3z^2-r^2},d_{xz},d_{x^2-y^2}\}$. We find that the $d$ partial PDOS is dominating
over $s$ and $p$ partial PDOS with the exception of the majority spin ($\uparrow$) PDOS above the Fermi levels. The obtained
results are in good agreement with Refs.\ \cite{heinze98,ferriani10tip}, where the full potential linearized augmented plane wave
(FLAPW) method was employed.

\begin{figure*}[ht!]
\includegraphics[width=0.78\textwidth,angle=0]{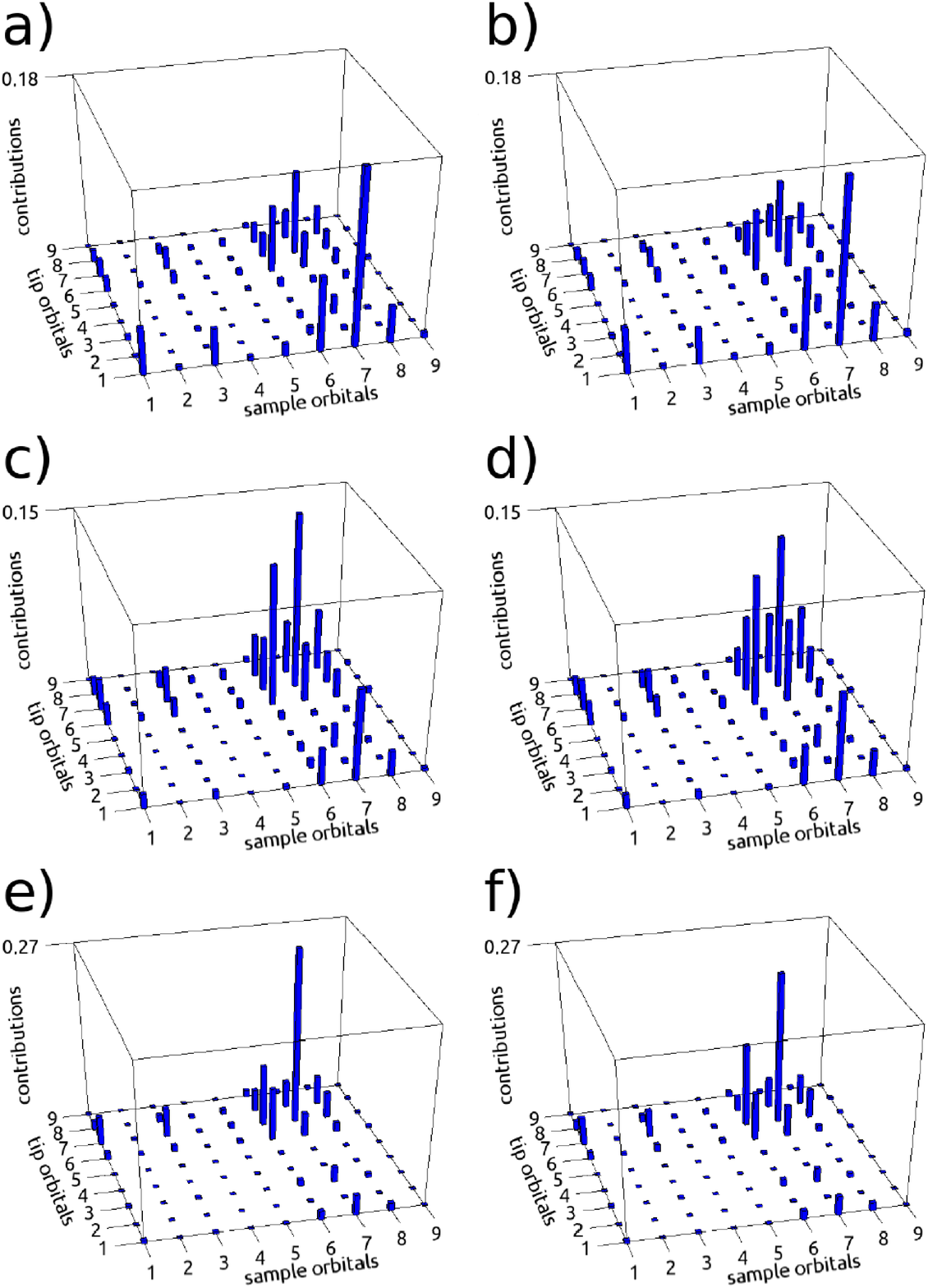}
\caption{\label{Fig3} (Color online) Histograms of the orbital-dependent relative current contributions
[$\tilde{I}_{TOT}^{\beta\gamma}$ in Eq.(\ref{Eq_i_bg})] using the iron tip placed $z=5$ \AA\;above
two Fe(110) surface positions [top (T) and hollow (H), see Figure \ref{Fig1}] at different bias voltages:
a) T position, $V$= -0.1 V; b) H position, $V$= -0.1 V;
c) T position, $V$= -1.0 V; d) H position, $V$= -1.0 V;
e) T position, $V$= -2.0 V; f) H position, $V$= -2.0 V.
The indices of the atomic orbitals (1-9) correspond to the order
$\{s,p_y,p_z,p_x,d_{xy},d_{yz},d_{3z^2-r^2},d_{xz},d_{x^2-y^2}\}$.
}
\end{figure*}

The relative contributions of all $\beta\leftrightarrow\gamma$ orbital-dependent transitions to the total tunneling current,
$\tilde{I}_{TOT}^{\beta\gamma}$, can be calculated according to Eq.(\ref{Eq_i_bg}). Figure \ref{Fig3} shows representative
histograms at different bias voltages and iron tip positions above the Fe(110) surface. We set the tip magnetization parallel to
the in-plane Fe(110) surface magnetic moment.
Parts a) and b) of Figure \ref{Fig3} correspond to the tip apex $z=5$ \AA\;above the surface top (T) and hollow (H) positions,
respectively (for T and H see Figure \ref{Fig1}), and the bias voltage is set to $V$= -0.1 V. We find that the tip $s$ (1)
orbital provides the dominating contributions combined with the sample $d_{yz}$ (6) and $d_{3z^2-r^2}$ (7) orbitals.
The $d_{3z^2-r^2}-d_{3z^2-r^2}$ (7-7) and $d_{yz}-d_{yz}$ (6-6) transitions also give sizeable contributions.
The main difference between the T and H tip positions is that above the Fe(110) hollow site the $d_{yz}-d_{3z^2-r^2}$
(6-7 and 7-6) contributions gain, while the $d_{3z^2-r^2}-d_{3z^2-r^2}$ (7-7) contribution loses weight, similarly to the
finding above the W(110) surface \cite{palotas12orb}. This is due to the different orientational overlap of the mentioned
tip and sample orbitals at the two tip positions, and it is not affected by the bias voltage. On the other hand,
at a larger negative bias $V$= -1.0 V [Figure \ref{Fig3}c) and \ref{Fig3}d)] we observe that the tip $s$ (1) orbital loses
and the $d_{yz}$ (6), $d_{3z^2-r^2}$ (7), and $d_{xz}$ (8) orbitals and their tip-sample combinations gain weight.
The largest contribution is now found for the $d_{3z^2-r^2}-d_{3z^2-r^2}$ (7-7) transition.
We find an enhancement of this effect at $V$= -2.0 V [Figure \ref{Fig3}e) and \ref{Fig3}f)], and here the tip $s$ (1) orbital
contributions are tiny. This bias-trend can be explained by the surface and tip partial charge PDOS in Figure \ref{Fig2}.
At large negative bias voltages the combination of the surface occupied $d$ partial PDOS with the tip unoccupied $d$ partial PDOS
clearly dominates over the tip $s$ partial contributions.
We find similar trends for a tip magnetization perpendicular to the surface, and also for positive bias voltages
(consider the combination of the surface unoccupied with the tip occupied $d$ partial charge PDOS in Figure \ref{Fig2}).

As demonstrated in Ref.\ \cite{palotas12orb}, the current difference between tip positions above the top (T) and hollow (H)
surface sites is indicative for the corrugation character of a constant-current STM image. The total ($TOT=TOPO+MAGN$) current
difference at tip-sample distance $z$ and bias voltage $V$ is defined as
\begin{equation}
\label{Eq_deltaI}
\Delta I_{TOT}(z,V)=I_{TOT}^T(z,V)-I_{TOT}^H(z,V).
\end{equation}
Positive $\Delta I_{TOT}$ corresponds to an STM image with normal corrugation where the atomic sites appear as protrusions,
and negative $\Delta I_{TOT}$ to an inverted STM image with anticorrugation and atomic sites appearing as depressions.
The $\Delta I_{TOT}(z,V)=0$ contour gives the $(z,V)$ combinations where the corrugation inversion occurs.

\begin{figure*}[h!]
\includegraphics[width=0.96\textwidth,angle=0]{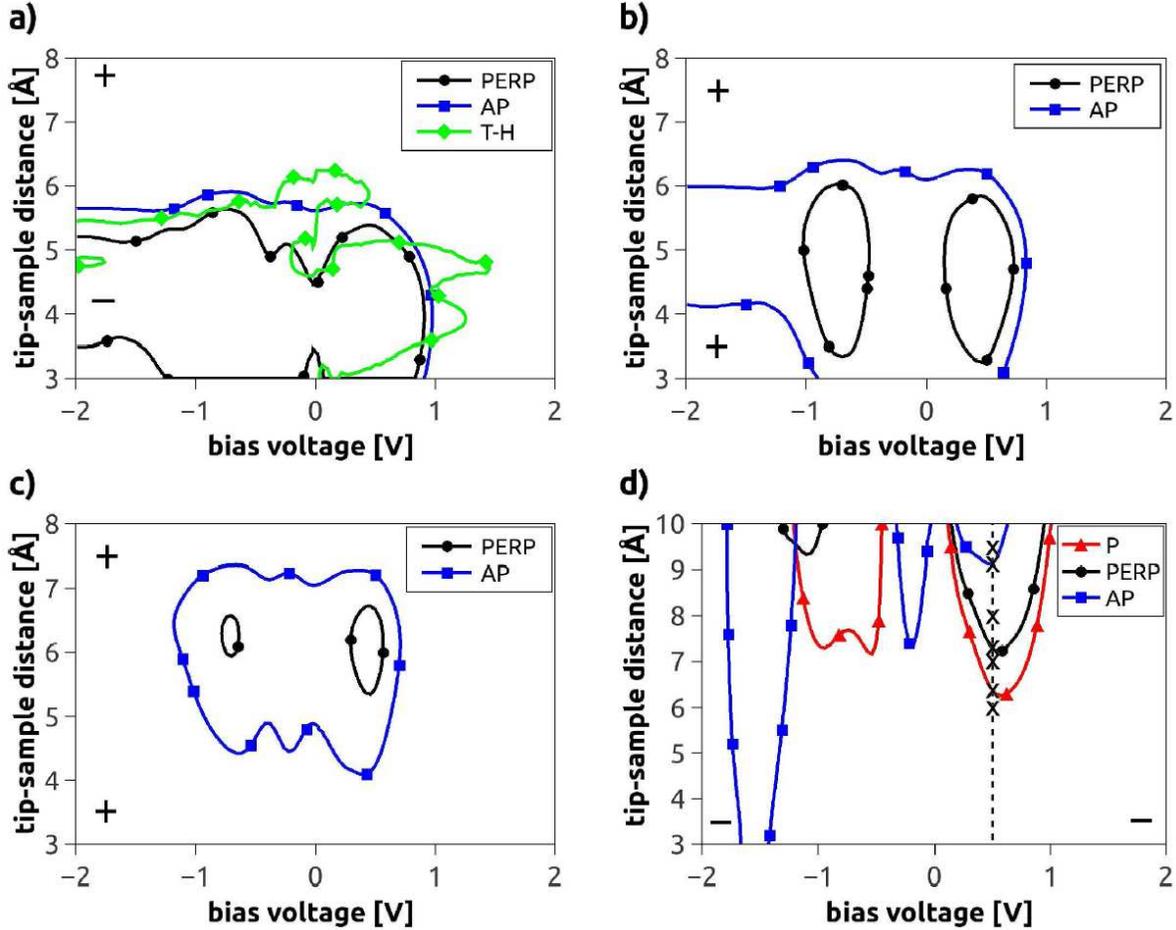}
\caption{\label{Fig4} (Color online) The $\Delta I_{TOT}(z,V)=I_{TOT}^T(z,V)-I_{TOT}^H(z,V)=0$ contours indicative for the
corrugation inversion [see Eq.(\ref{Eq_deltaI}), and its meaning in the text] calculated with four tip models above the
Fe(110) surface:
a) $s$-tip, b) $p_z$-tip, c) $d_{3z^2-r^2}$-tip, and d) Fe(001)-tip. The tip magnetization directions are explicitly shown
with respect to the Fe(110) surface magnetic moment: parallel (P), perpendicular (PERP), and antiparallel (AP). For comparison,
the $\Delta I_{TOT}(z,V)=0$ curves are shown in Fig.\ \ref{Fig4}a) obtained by using the Tersoff-Hamann (T-H) approach.
The sign of $\Delta I_{TOT}$ ($+$ or $-$) is shown in selected $(z,V)$ sections, and crossing the $\Delta I_{TOT}=0$ curve
always means inversion of the sign. Note that positive $\Delta I_{TOT}$ corresponds to normal, and negative to
inverted atomic contrast in the SP-STM images, see also Fig.\ \ref{Fig5}. In Fig.\ \ref{Fig4}d) seven crosses at 0.5 V
mark the apparent heights of the Fe atom on the constant-current contours shown in Fig.\ \ref{Fig5}.
}
\end{figure*}

Figure \ref{Fig4} shows such zero current difference contours calculated with four different tip models and three tip
magnetization orientations [parallel (P), perpendicular (PERP), and antiparallel (AP) with respect to the Fe(110) surface
magnetic moment] in the [-2 V, +2 V] bias voltage range. We consider the [3 \AA, 8 \AA] tip-sample distance
regime in Figs.\ \ref{Fig4}a), \ref{Fig4}b), and \ref{Fig4}c) for the ideal tips with $s$, $p_z$, and $d_{3z^2-r^2}$ orbital
character, respectively, and the [3 \AA, 10 \AA] range in Fig.\ \ref{Fig4}d) for the iron tip.
Note that the validity of our tunneling model is restricted to about $z>3.5$ \AA.

Let us first focus on the zero current difference contours obtained
by the $s$-tip in Fig.\ \ref{Fig4}a). The contrast inversion of the Fe(110) surface with an $s$-tip at a fixed $z=4.5$ \AA\;
tip-sample distance was already studied with a different theoretical approach by Heinze et al.\ in Ref.\ \cite{heinze98}.
Comparing these results with ours we find good qualitative agreement: The P tip magnetization does not show any contrast
inversion in the studied $(z,V)$ range. This corresponds to the majority spin$+$ data of Fig.\ 15 in Ref.\ \cite{heinze98}.
On the other hand, the PERP and AP tip magnetizations result in $\Delta I_{TOT}(z,V)=0$ curves shifted away from each other,
and the direction of the shift qualitatively agrees with the findings of Ref.\ \cite{heinze98}. Note, however, that the
corrugation inversions at $z=4.5$ \AA\;are found at different bias voltages: 0.87 V (our model, PERP) vs.\ 0.4 V
(Total, Fig.\ 15 in Ref.\ \cite{heinze98}), and 0.94 V (our model, AP) vs.\ 0.7 V
(minority spin$-$, Fig.\ 15 in Ref.\ \cite{heinze98}). This difference can be attributed to the nonequal lattice constants
used and the different theoretical approaches.

As a further test of the reliability of our tunneling model, we compared the $\Delta I_{TOT}(z,V)=0$ contour
using the PERP tip magnetization with that obtained by the Tersoff-Hamann method, the curve denoted by T-H in Fig.\ \ref{Fig4}a).
We find that our model reproduces the $(z,V)$ region with the inverse corrugation denoted by the '$-$' sign qualitatively well.
Quantitative agreement is not present due to the approximations in the 3D WKB method \cite{palotas13fop}.
Note that the PERP tip magnetization corresponds to an equal 0.5-0.5 weighting of the majority and minority spin channels
contributing to the tunneling and, thus, to a zero spin-polarized ($MAGN$) contribution of the current ($I_{MAGN}=0$,
$I_{TOT}=I_{TOPO}$), and solely orbital-dependent effects play a role in the tunneling without spin-polarization effects.

Altogether, the results obtained by the $s$-tip indicate that the Fe(110) surface always appears to be normally corrugated
taking a P tip magnetization with respect to the surface magnetic moment, whereas changing the tip magnetization to PERP and AP
result in anticorrugation below about 1 V bias and 6 \AA\;tip-sample distance.

To understand these results, we give the following interpretation: It is known that there is a competition
between the tunneling contributions from $m=0$ ($s$, $p_z$, $d_{3z^2-r^2}$) and $m\ne 0$ orbitals in the formation of the
corrugation character of an STM image \cite{chen92,palotas12orb}. Generally, $m=0$ states prefer normal corrugation, whereas
$m\ne 0$ states prefer anticorrugation.
Based on our model, we suggest that the corrugation inversion can be understood as an interplay of the real-space orbital shapes
involved in the tunneling and their corresponding energy-dependent partial PDOS. For the PERP [Fig.\ \ref{Fig2}a)] and AP
[minority spin ($\downarrow$) in Fig.\ \ref{Fig2}c)] cases the $d$ partial PDOS of the Fe(110) surface is dominating over the $s$
and $p$ partial PDOS in the whole energy range. Therefore, considering only $d$ orbital shapes of the surface, the leading current
contribution is expected from the $d_{3z^2-r^2}$ orbital of the underlying Fe atom when the tip is placed above the surface T
position, and from the $d_{3z^2-r^2}$, $d_{xz,yz}$ orbitals of the four nearest-neighbor surface Fe atoms when the tip is above
the surface H position. Note that the weight of $d_{yz}$ is larger than of $d_{xz}$ due to the geometry of the (110) surface, see
Fig.\ \ref{Fig1}. By changing the tip-sample distance there is a competition between the contributions of these $d$ states due to
the orbital-dependent transmission in Eq.\ \ref{Eq_Transmission_decomp}, and there is a range ($z<6$ \AA) where $I_{TOT}^H$
dominates over $I_{TOT}^T$, thus anticorrugation is obtained. At large distances the $d_{xz,yz}$ orbitals lose and the
$d_{3z^2-r^2}$ orbital gains importance, and normal corrugation is found. This simple orbital-picture has to be combined with the
energy-dependence of the PDOS of the contributing spin channels for an accurate interpretation. As can be seen in
Figs.\ \ref{Fig2}a) and \ref{Fig2}c) ($\downarrow$), there is a large $d_{3z^2-r^2}$-type PDOS peak at about 1.5 eV above the
Fermi level. The existence of this PDOS feature results in the disappearance of the anticorrugated region above around 1 V.
Note that although the $d_{xz}$-type PDOS is comparable in size with the $d_{3z^2-r^2}$-type PDOS in this energy range, the
weighted contribution of $d_{xz}$ orbitals is much smaller due to the orbital shapes and the tip-sample geometry. For the P tip
magnetization, below the Fermi level the orbital-shape-weighted contribution of $d_{3z^2-r^2}$ is dominating compared to
$d_{xz,yz}$, for the PDOS see Fig.\ \ref{Fig2}c) ($\uparrow$). Above the Fermi level the importance of the $d$ states decreases
rapidly, and concomitantly $m=0$ $s$ and $p_z$ states gain more importance, resulting in normal corrugation in the full energy
and tip-sample distance range.

Comparing Figs.\ \ref{Fig4}a), \ref{Fig4}b), and \ref{Fig4}c) we find that the anticorrugation regime is shifted to larger
tip-sample distances following the order $s$, $p_z$, and $d_{3z^2-r^2}$, similarly as found for the W(110) surface
\cite{palotas12orb}. This is due to the increasingly localized character of these tip orbitals in the $z$-direction.
It means that the dominating current contributions originate from less and less surface atoms that are
located directly below the tip position. Considering the orbital-dependent transmission in Eq.\ \ref{Eq_Transmission_decomp}
and these $z$-localized ($m=0$) tip orbitals, it follows naturally that the maximal effect of the surface $d_{xz,yz}$ orbitals
preferring anticorrugation is found at larger tip-sample distances.
On the other hand, there are certain energy regimes where this orbital overlap effect results in the occurrence of
normal corrugation since the complex competition between $d_{xz,yz}$ and $d_{3z^2-r^2}$ surface orbitals weighted with their
partial PDOS is won by the corrugating $d_{3z^2-r^2}$ state.
For all considered ideal magnetic tips the P tip magnetization corresponding to the majority spin channel does not show any
contrast inversion in the studied $(z,V)$ range. Taking the PERP and AP tip magnetization directions we expect
normal corrugation above approximately 1.0 V ($s$-tip), 0.9 V ($p_z$-tip), and 0.8 V ($d_{3z^2-r^2}$-tip) irrespective of the
tip-sample distance. The reasons are outlined above, and see the discussion for the $s$-tip.
The series of Figs.\ \ref{Fig4}a), \ref{Fig4}b), and \ref{Fig4}c) demonstrate
the orbital-dependent tunneling and tip orbital effects on the corrugation inversion.

Taking the iron tip having all nine orbital characters with weighted energy-dependent PDOS, it is clearly seen in
Fig.\ \ref{Fig4}d) that the $\Delta I(z,V)=0$ contours are considerably affected by the tip electronic structure.
We find that all considered tip magnetization directions result in the appearance of corrugation inversions in the studied
bias voltage and tip-sample distance range. Close to the surface anticorrugation is observed for the same reason
as discussed for the $s$-tip, with the exception of the approximate [-1.7 V, -1.4 V] range where the AP tip magnetization shows
normal corrugation and no inversion at all. The reasons for this finding are the observed $d_{3z^2-r^2}$ peaks in the surface
majority spin ($\uparrow$) PDOS at around 1 eV and 1.5 eV below the sample Fermi energy and the tip minority spin ($\downarrow$)
PDOS at around 1.5 eV above the tip Fermi level, see Figs.\ \ref{Fig2}c) and \ref{Fig2}d). Such a combination of electronic
structures gives a robust normal corrugation in the given bias range. By moving the tip away from the surface, contrast inversions
are indicated by the $\Delta I=0$ contours at large tip-sample distances, i.e., the anticorrugation ($\Delta I<0$) changes to
normal corrugation ($\Delta I>0$). It is clearly seen that these inversion contours vary considerably depending on the tip
magnetization orientation. This is due to the complex combined effect of the orbital-dependent tunneling and
spin-polarization originating from the electronic structures of the sample surface and the tip.
For instance, the minority spin ($\downarrow$) $d_{xy}$ PDOS peak of the tip located at 0.5 eV below its Fermi
level in Fig.\ \ref{Fig2}d) is mainly responsible for the set of observed inversions at around 0.5 V bias. The $d_{xy}$ tip
orbital prefers normal corrugation in the studied tip-sample distance range due to its $xz$ and $yz$ nodal planes since above the
H position the anticorrugating $d_{xz,yz}$ surface orbitals do not contribute to the current, and $I_{TOT}^T$ is expected to be
larger than $I_{TOT}^H$ \cite{palotas12orb}. Considering other orbital contributions weighted with their partial PDOS, however,
results in anticorrugation below about $z=6$ \AA. In effect, we point out that the complex interplay of the real-space orbital
shapes and the energy-dependent PDOS of the contributing spin channels of both the surface and the tip results in the observed
corrugation inversion maps in Fig.\ \ref{Fig4}. The above findings demonstrate the tunability of the height of the
observable atomic contrast inversion depending on the spin-polarized tunneling via the tip magnetization orientation.

\begin{figure*}[h!]
\includegraphics[width=1.0\textwidth,angle=0]{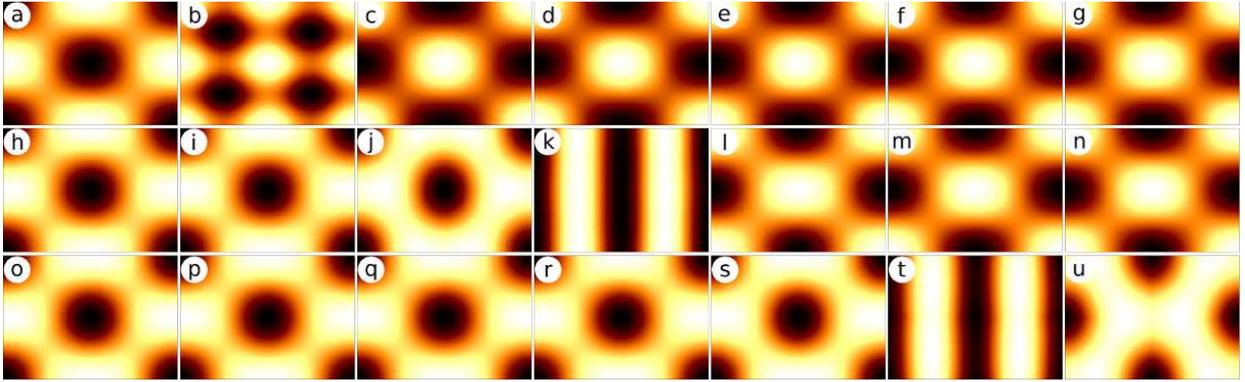}
\caption{\label{Fig5} (Color online) Simulated constant-current SP-STM images of the Fe(110) surface at $V$= 0.5 V
bias voltage and three magnetic orientations of the iron tip: parallel (a-g), perpendicular (h-n), and antiparallel (o-u) to the
Fe(110) surface magnetic moment. The apparent heights of the Fe atom denoted by crosses in Fig.\ \ref{Fig4}d) are the same in each
column:
a), h), o) 6.00 \AA;
b), i), p) 6.37 \AA;
c), j), q) 7.00 \AA;
d), k), r) 7.28 \AA;
e), l), s) 8.00 \AA;
f), m), t) 9.12 \AA;
g), n), u) 9.50 \AA.
The current values are as follows:
a) 509 pA, b) 234 pA, c)  62 pA, d)  35 pA, e) 7.6 pA, f) 0.7 pA, g) 0.3 pA,
h) 468 pA, i) 215 pA, j)  58 pA, k)  32 pA, l) 7.1 pA, m) 0.7 pA, n) 0.3 pA,
o) 427 pA, p) 197 pA, q)  53 pA, r)  29 pA, s) 6.5 pA, t) 0.6 pA, u) 0.3 pA.
The SP-STM scan area corresponds to the rectangle shown in Figure \ref{Fig1}. Light and dark areas denote larger and smaller
apparent heights, respectively.
}
\end{figure*}

In order to show the dependence of the atomic contrast inversion on the tip-sample distance and on the
tip magnetization orientation more apparently, constant-current SP-STM images are simulated.
We choose the iron tip and the bias voltage of $V=$ 0.5 V.
The total tunneling current ($I_{TOT}$) is calculated in sheets of 0.01 \AA\;vertical width centered at seven selected
tip heights between 6 and 9.5 \AA\;marked by crosses in Fig.\ \ref{Fig4}d) above the rectangular scan area shown in
Figure \ref{Fig1}. The lateral resolution is 0.04 \AA\;by taking $100 \times 70$ points in each sheet.
The constant-current contours are extracted following the method described in Ref.\ \cite{palotas11stm}.
Figure \ref{Fig5} shows the simulated constant-current SP-STM images of the Fe(110) surface at seven apparent heights
of the surface Fe atom (seven columns) and three magnetic orientations of the iron tip (three rows): parallel (P, first row: a-g),
perpendicular (PERP, second row: h-n), and antiparallel (AP, third row: o-u) to the Fe(110) surface magnetic moment.
The apparent heights of the Fe atom and the current values of the contours are given in the caption of Figure \ref{Fig5}.
Going from the left to the right in Figure \ref{Fig5}, the current value is decreasing with a concomitant movement of the tip
farther from the surface. It can clearly be seen that the contrast inversion occurs at different tip heights depending
on the tip magnetization orientation: P, 6.37 \AA\;(b); PERP, 7.28 \AA\;(k); AP, 9.12 \AA\;(t), see also Fig.\ \ref{Fig4}d).
We find that the SP-STM image of the contrast inversion with P tip magnetization orientation (Fig.\ \ref{Fig5}b) is markedly
different from the striped images in the PERP and AP case (Figs.\ \ref{Fig5}k and \ref{Fig5}t, respectively).
Such striped STM images of the contrast inversion were also reported above the nonmagnetic W(110) surface \cite{palotas12orb}.
Note, however, that in experiments the atomic resolution is lost close to the inversion \cite{heinze98}.
Below the inversion we observe anticorrugation (Figs.\ \ref{Fig5}a, h-j, o-s), i.e., the surface atoms appear as
depressions, and above the inversion normal corrugation is obtained (Figs.\ \ref{Fig5}c-g, l-n, u), in accordance
with Fig.\ \ref{Fig4}d).

\section{Conclusions}
\label{sec_conc}

We extended the orbital-dependent electron tunneling model implemented within the three-dimensional (3D) Wentzel-Kramers-Brillouin
(WKB) atom-superposition approach to simulate spin-polarized scanning tunneling microscopy (SP-STM) above magnetic surfaces.
Applying our method, we analyzed the bias-dependence of the orbital contributions to the tunneling current above the Fe(110)
surface, and found a shift of the relevant tip $s$ contributions close to zero bias toward $d-d$ tunneling at higher bias.
We showed that spin-polarized tunneling has a considerable effect on the tip-sample distance where atomic contrast inversion
occurs, and the tip magnetization direction and tip orbital composition play a crucial role as well.
Taking an $s$-tip, our findings showed qualitative agreement with the Tersoff-Hamann method and with Ref.\ \cite{heinze98}
concerning the corrugation character of the Fe(110) surface. We explained our results based on the
complex interplay of the real-space orbital shapes involved in the tunneling and the energy-dependent
orbital-decomposed PDOS of the sample and the tip. Finally, we demonstrated the contrast inversion by simulating SP-STM images.

\section{Acknowledgments}

Financial support of the Magyary Foundation, EEA and Norway Grants, the Hungarian Scientific Research Fund (OTKA PD83353, K77771),
the Bolyai Research Grant of the Hungarian Academy of Sciences, and the New Sz\'echenyi Plan of Hungary
(Project ID: T\'AMOP-4.2.2.B-10/1--2010-0009) is gratefully acknowledged. Usage
of the computing facilities of the Wigner Research Centre for Physics, and the BME HPC Cluster is kindly acknowledged.


\begin{thebibliography}{99}

\expandafter\ifx\csname natexlab\endcsname\relax\def\natexlab#1{#1}\fi
\expandafter\ifx\csname bibnamefont\endcsname\relax
  \def\bibnamefont#1{#1}\fi
\expandafter\ifx\csname bibfnamefont\endcsname\relax
  \def\bibfnamefont#1{#1}\fi
\expandafter\ifx\csname citenamefont\endcsname\relax
  \def\citenamefont#1{#1}\fi
\expandafter\ifx\csname url\endcsname\relax
  \def\url#1{\texttt{#1}}\fi
\expandafter\ifx\csname urlprefix\endcsname\relax\def\urlprefix{URL }\fi
\providecommand{\bibinfo}[2]{#2}
\providecommand{\eprint}[2][]{\url{#2}}

\bibitem{wiesendanger09review}
\bibinfo{author}{\bibfnamefont{R.}~\bibnamefont{Wiesendanger}},
\bibinfo{title}{\bibfnamefont{Spin mapping at the nanoscale and atomic scale}},
\bibinfo{journal}{Rev. Mod. Phys.} \bibinfo{volume}{81}
(\bibinfo{year}{2009}) \bibinfo{pages}{1495-1550}.

\bibitem{hofer03rmp}
\bibinfo{author}{\bibfnamefont{W.A.}~\bibnamefont{Hofer}},
\bibinfo{author}{\bibfnamefont{A.S.}~\bibnamefont{Foster}},
\bibinfo{author}{\bibfnamefont{A.L.}~\bibnamefont{Shluger}},
\bibinfo{title}{\bibfnamefont{Theories of scanning probe microscopes at the atomic scale}},
\bibinfo{journal}{Rev. Mod. Phys.} \bibinfo{volume}{75}
(\bibinfo{year}{2003}) \bibinfo{pages}{1287-1331}.

\bibitem{hofer03pssci}
\bibinfo{author}{\bibfnamefont{W.A.}~\bibnamefont{Hofer}},
\bibinfo{title}{\bibfnamefont{Challenges and errors: interpreting high resolution images in scanning tunneling microscopy}},
\bibinfo{journal}{Prog. Surf. Sci.} \bibinfo{volume}{71}
(\bibinfo{year}{2003}) \bibinfo{pages}{147-183}.

\bibitem{palotas12orb}
\bibinfo{author}{\bibfnamefont{K.}~\bibnamefont{Palot\'as}},
\bibinfo{author}{\bibfnamefont{G.}~\bibnamefont{M\'andi}},
\bibinfo{author}{\bibfnamefont{L.}~\bibnamefont{Szunyogh}},
\bibinfo{title}{\bibfnamefont{Orbital-dependent electron tunneling within the atom superposition approach: Theory and application to W(110)}},
\bibinfo{journal}{Phys. Rev. B} \bibinfo{volume}{86}
(\bibinfo{year}{2012}) \bibinfo{pages}{235415}.

\bibitem{mingo96}
\bibinfo{author}{\bibfnamefont{N.} \bibnamefont{Mingo}},
\bibinfo{author}{\bibfnamefont{L.} \bibnamefont{Jurczyszyn}},
\bibinfo{author}{\bibfnamefont{F.J.} \bibnamefont{Garcia-Vidal}},
\bibinfo{author}{\bibfnamefont{R.} \bibnamefont{Saiz-Pardo}},
\bibinfo{author}{\bibfnamefont{P.L.} \bibnamefont{de Andres}},
\bibinfo{author}{\bibfnamefont{F.} \bibnamefont{Flores}},
\bibinfo{author}{\bibfnamefont{S.Y.} \bibnamefont{Wu}},
\bibinfo{author}{\bibfnamefont{W.} \bibnamefont{More}},
\bibinfo{title}{\bibfnamefont{Theory of the scanning tunneling microscope: Xe on Ni and Al}},
\bibinfo{journal}{Phys. Rev. B} \bibinfo{volume}{54}
(\bibinfo{year}{1996}) \bibinfo{pages}{2225-2235}.

\bibitem{heinze98}
\bibinfo{author}{\bibfnamefont{S.}~\bibnamefont{Heinze}},
\bibinfo{author}{\bibfnamefont{S.}~\bibnamefont{Bl\"ugel}},
\bibinfo{author}{\bibfnamefont{R.}~\bibnamefont{Pascal}},
\bibinfo{author}{\bibfnamefont{M.}~\bibnamefont{Bode}},
\bibinfo{author}{\bibfnamefont{R.}~\bibnamefont{Wiesendanger}},
\bibinfo{title}{\bibfnamefont{Prediction of bias-voltage-dependent corrugation reversal for STM images of bcc (110) surfaces: W(110), Ta(110), and Fe(110)}},
\bibinfo{journal}{Phys. Rev. B} \bibinfo{volume}{58}
(\bibinfo{year}{1998}) \bibinfo{pages}{16432-16445}.

\bibitem{ondracek12}
\bibinfo{author}{\bibfnamefont{M.} \bibnamefont{Ondr\'a\v{c}ek}},
\bibinfo{author}{\bibfnamefont{C.} \bibnamefont{Gonz\'alez}},
\bibinfo{author}{\bibfnamefont{P.} \bibnamefont{Jel\'inek}},
\bibinfo{title}{\bibfnamefont{Reversal of atomic contrast in scanning probe microscopy on (111) metal surfaces}},
\bibinfo{journal}{J. Phys. Condens. Matter} \bibinfo{volume}{24}
(\bibinfo{year}{2012}) \bibinfo{pages}{084003}.

\bibitem{chen92}
\bibinfo{author}{\bibfnamefont{C.J.} \bibnamefont{Chen}},
\bibinfo{title}{\bibfnamefont{Effects of m$\ne$0 tip states in scanning tunneling microscopy: The explanations of corrugation reversal}},
\bibinfo{journal}{Phys. Rev. Lett.} \bibinfo{volume}{69}
(\bibinfo{year}{1992}) \bibinfo{pages}{1656-1659}.

\bibitem{heinze99}
\bibinfo{author}{\bibfnamefont{S.} \bibnamefont{Heinze}},
\bibinfo{author}{\bibfnamefont{X.} \bibnamefont{Nie}},
\bibinfo{author}{\bibfnamefont{S.} \bibnamefont{Bl\"ugel}},
\bibinfo{author}{\bibfnamefont{M.} \bibnamefont{Weinert}},
\bibinfo{title}{\bibfnamefont{Electric-field-induced changes in scanning tunneling microscopy images of metal surfaces}},
\bibinfo{journal}{Chem. Phys. Lett.} \bibinfo{volume}{315}
(\bibinfo{year}{1999}) \bibinfo{pages}{167-172}.

\bibitem{yang06}
\bibinfo{author}{\bibfnamefont{R.}~\bibnamefont{Yang}},
\bibinfo{author}{\bibfnamefont{H.}~\bibnamefont{Yang}},
\bibinfo{author}{\bibfnamefont{A.R.}~\bibnamefont{Smith}},
\bibinfo{author}{\bibfnamefont{A.}~\bibnamefont{Dick}},
\bibinfo{author}{\bibfnamefont{J.}~\bibnamefont{Neugebauer}},
\bibinfo{title}{\bibfnamefont{Energy-dependent contrast in atomic-scale spin-polarized scanning tunneling microscopy of Mn$_3$N$_2$(010): Experiment and first-principles theory}},
\bibinfo{journal}{Phys. Rev. B} \bibinfo{volume}{74}
(\bibinfo{year}{2006}) \bibinfo{pages}{115409}.

\bibitem{palotas11stm}
\bibinfo{author}{\bibfnamefont{K.}~\bibnamefont{Palot\'as}},
\bibinfo{author}{\bibfnamefont{W.A.}~\bibnamefont{Hofer}},
\bibinfo{author}{\bibfnamefont{L.}~\bibnamefont{Szunyogh}},
\bibinfo{title}{\bibfnamefont{Simulation of spin-polarized scanning tunneling microscopy on complex magnetic surfaces: Case of a Cr monolayer on Ag(111)}},
\bibinfo{journal}{Phys. Rev. B} \bibinfo{volume}{84}
(\bibinfo{year}{2011}) \bibinfo{pages}{174428}.

\bibitem{palotas13contrast}
\bibinfo{author}{\bibfnamefont{K.}~\bibnamefont{Palot\'as}},
\bibinfo{title}{\bibfnamefont{Prediction of the bias voltage dependent magnetic contrast in spin-polarized scanning tunneling microscopy}},
\bibinfo{journal}{Phys. Rev. B} \bibinfo{volume}{87}
(\bibinfo{year}{2013}) \bibinfo{pages}{024417}.

\bibitem{hofer08tipH}
\bibinfo{author}{\bibfnamefont{W.A.}~\bibnamefont{Hofer}},
\bibinfo{author}{\bibfnamefont{K.}~\bibnamefont{Palot\'as}},
\bibinfo{author}{\bibfnamefont{S.}~\bibnamefont{Rusponi}},
\bibinfo{author}{\bibfnamefont{T.}~\bibnamefont{Cren}},
\bibinfo{author}{\bibfnamefont{H.}~\bibnamefont{Brune}},
\bibinfo{title}{\bibfnamefont{Role of hydrogen in giant spin polarization observed on magnetic nanostructures}},
\bibinfo{journal}{Phys. Rev. Lett.} \bibinfo{volume}{100}
(\bibinfo{year}{2008}) \bibinfo{pages}{026806}.

\bibitem{wortmann01}
\bibinfo{author}{\bibfnamefont{D.}~\bibnamefont{Wortmann}},
\bibinfo{author}{\bibfnamefont{S.}~\bibnamefont{Heinze}},
\bibinfo{author}{\bibfnamefont{P.}~\bibnamefont{Kurz}},
\bibinfo{author}{\bibfnamefont{G.}~\bibnamefont{Bihlmayer}},
\bibinfo{author}{\bibfnamefont{S.}~\bibnamefont{Bl\"ugel}},
\bibinfo{title}{\bibfnamefont{Resolving complex atomic-scale spin structures by spin-polarized scanning tunneling microscopy}},
\bibinfo{journal}{Phys. Rev. Lett.} \bibinfo{volume}{86}
(\bibinfo{year}{2001}) \bibinfo{pages}{4132-4135}.

\bibitem{hofer03}
\bibinfo{author}{\bibfnamefont{W.A.}~\bibnamefont{Hofer}},
\bibinfo{author}{\bibfnamefont{A.J.}~\bibnamefont{Fisher}},
\bibinfo{title}{\bibfnamefont{Simulation of spin-resolved scanning tunneling microscopy: influence of the magnetization of surface and tip}},
\bibinfo{journal}{J. Magn. Magn. Mater.} \bibinfo{volume}{267}
(\bibinfo{year}{2003}) \bibinfo{pages}{139-151}.

\bibitem{smith04}
\bibinfo{author}{\bibfnamefont{A.R.}~\bibnamefont{Smith}},
\bibinfo{author}{\bibfnamefont{R.}~\bibnamefont{Yang}},
\bibinfo{author}{\bibfnamefont{H.}~\bibnamefont{Yang}},
\bibinfo{author}{\bibfnamefont{W.R.L.}~\bibnamefont{Lambrecht}},
\bibinfo{author}{\bibfnamefont{A.}~\bibnamefont{Dick}},
\bibinfo{author}{\bibfnamefont{J.}~\bibnamefont{Neugebauer}},
\bibinfo{title}{\bibfnamefont{Aspects of spin-polarized scanning tunneling microscopy at the atomic scale: experiment, theory, and simulation}},
\bibinfo{journal}{Surf. Sci.} \bibinfo{volume}{561}
(\bibinfo{year}{2004}) \bibinfo{pages}{154-170}.

\bibitem{heinze06}
\bibinfo{author}{\bibfnamefont{S.}~\bibnamefont{Heinze}},
\bibinfo{title}{\bibfnamefont{Simulation of spin-polarized scanning tunneling microscopy images of nanoscale non-collinear magnetic structures}},
\bibinfo{journal}{Appl. Phys. A} \bibinfo{volume}{85}
(\bibinfo{year}{2006}) \bibinfo{pages}{407-414}.

\bibitem{tersoff85}
\bibinfo{author}{\bibfnamefont{J.}~\bibnamefont{Tersoff}},
\bibinfo{author}{\bibfnamefont{D.R.}~\bibnamefont{Hamann}},
\bibinfo{title}{\bibfnamefont{Theory of the scanning tunneling microscope}},
\bibinfo{journal}{Phys. Rev. B} \bibinfo{volume}{31}
(\bibinfo{year}{1985}) \bibinfo{pages}{805-813}.

\bibitem{yang02}
\bibinfo{author}{\bibfnamefont{H.}~\bibnamefont{Yang}},
\bibinfo{author}{\bibfnamefont{A.R.}~\bibnamefont{Smith}},
\bibinfo{author}{\bibfnamefont{M.}~\bibnamefont{Prikhodko}},
\bibinfo{author}{\bibfnamefont{W.R.L.}~\bibnamefont{Lambrecht}},
\bibinfo{title}{\bibfnamefont{Atomic-scale spin-polarized scanning tunneling microscopy applied to Mn$_3$N$_2$(010)}},
\bibinfo{journal}{Phys. Rev. Lett.} \bibinfo{volume}{89}
(\bibinfo{year}{2002}) \bibinfo{pages}{226101}.

\bibitem{tersoff83}
\bibinfo{author}{\bibfnamefont{J.}~\bibnamefont{Tersoff}},
\bibinfo{author}{\bibfnamefont{D.R.}~\bibnamefont{Hamann}},
\bibinfo{title}{\bibfnamefont{Theory and application for the scanning tunneling microscope}},
\bibinfo{journal}{Phys. Rev. Lett.} \bibinfo{volume}{50}
(\bibinfo{year}{1983}) \bibinfo{pages}{1998-2001}.

\bibitem{bardeen61}
\bibinfo{author}{\bibfnamefont{J.}~\bibnamefont{Bardeen}},
\bibinfo{title}{\bibfnamefont{Tunnelling from a many-particle point of view}},
\bibinfo{journal}{Phys. Rev. Lett.} \bibinfo{volume}{6}
(\bibinfo{year}{1961}) \bibinfo{pages}{57-59}.

\bibitem{palotas05}
\bibinfo{author}{\bibfnamefont{K.}~\bibnamefont{Palot\'as}},
\bibinfo{author}{\bibfnamefont{W.A.}~\bibnamefont{Hofer}},
\bibinfo{title}{\bibfnamefont{Multiple scattering in a vacuum barrier obtained from real-space wavefunctions}},
\bibinfo{journal}{J. Phys. Condens. Matter} \bibinfo{volume}{17}
(\bibinfo{year}{2005}) \bibinfo{pages}{2705-2713}.

\bibitem{palotas13fop}
\bibinfo{author}{\bibfnamefont{K.}~\bibnamefont{Palot\'as}},
\bibinfo{author}{\bibfnamefont{G.}~\bibnamefont{M\'andi}},
\bibinfo{author}{\bibfnamefont{W.A.}~\bibnamefont{Hofer}},
\bibinfo{title}{\bibfnamefont{Three-dimensional Wentzel-Kramers-Brillouin approach for the simulation of scanning tunneling microscopy and spectroscopy}},
\bibinfo{journal}{Front. Phys.} \bibinfo{volume}{}
(\bibinfo{year}{2013}) \bibinfo{pages}{DOI: 10.1007/s11467-013-0354-4}.

\bibitem{palotas11sts}
\bibinfo{author}{\bibfnamefont{K.}~\bibnamefont{Palot\'as}},
\bibinfo{author}{\bibfnamefont{W.A.}~\bibnamefont{Hofer}},
\bibinfo{author}{\bibfnamefont{L.}~\bibnamefont{Szunyogh}},
\bibinfo{title}{\bibfnamefont{Theoretical study of the role of the tip in enhancing the sensitivity of differential conductance tunneling spectroscopy on magnetic surfaces}},
\bibinfo{journal}{Phys. Rev. B} \bibinfo{volume}{83}
(\bibinfo{year}{2011}) \bibinfo{pages}{214410}.

\bibitem{palotas12sts}
\bibinfo{author}{\bibfnamefont{K.}~\bibnamefont{Palot\'as}},
\bibinfo{author}{\bibfnamefont{W.A.}~\bibnamefont{Hofer}},
\bibinfo{author}{\bibfnamefont{L.}~\bibnamefont{Szunyogh}},
\bibinfo{title}{\bibfnamefont{Simulation of spin-polarized scanning tunneling spectroscopy on complex magnetic surfaces: Case of a Cr monolayer on Ag(111)}},
\bibinfo{journal}{Phys. Rev. B} \bibinfo{volume}{85}
(\bibinfo{year}{2012}) \bibinfo{pages}{205427}.

\bibitem{mandi13tiprot}
\bibinfo{author}{\bibfnamefont{G.}~\bibnamefont{M\'andi}},
\bibinfo{author}{\bibfnamefont{N.}~\bibnamefont{Nagy}},
\bibinfo{author}{\bibfnamefont{K.}~\bibnamefont{Palot\'as}},
\bibinfo{title}{\bibfnamefont{Arbitrary tip orientation in STM simulations: 3D WKB theory and application to W(110)}},
\bibinfo{journal}{J. Phys. Condens. Matter} \bibinfo{volume}{25}
(\bibinfo{year}{2013}) \bibinfo{pages}{445009}.

\bibitem{VASP2}
\bibinfo{author}{\bibfnamefont{G.}~\bibnamefont{Kresse}},
\bibinfo{author}{\bibfnamefont{J.}~\bibnamefont{Furthm\"uller}},
\bibinfo{title}{\bibfnamefont{Efficiency of ab-initio total energy calculations for metals and semiconductors using a plane-wave basis set}},
\bibinfo{journal}{Comput. Mater. Sci.} \bibinfo{volume}{6}
(\bibinfo{year}{1996}) \bibinfo{pages}{15-50}.

\bibitem{VASP3}
\bibinfo{author}{\bibfnamefont{G.}~\bibnamefont{Kresse}},
\bibinfo{author}{\bibfnamefont{J.}~\bibnamefont{Furthm\"uller}},
\bibinfo{title}{\bibfnamefont{Efficient iterative schemes for ab initio total-energy calculations using a plane-wave basis set}},
\bibinfo{journal}{Phys. Rev. B} \bibinfo{volume}{54}
(\bibinfo{year}{1996}) \bibinfo{pages}{11169-11186}.

\bibitem{hafner08}
\bibinfo{author}{\bibfnamefont{J.}~\bibnamefont{Hafner}},
\bibinfo{title}{\bibfnamefont{Ab-initio simulations of materials using VASP: Density-functional theory and beyond}},
\bibinfo{journal}{J. Comput. Chem.} \bibinfo{volume}{29}
(\bibinfo{year}{2008}) \bibinfo{pages}{2044-2078}.

\bibitem{kresse99}
\bibinfo{author}{\bibfnamefont{G.}~\bibnamefont{Kresse}},
\bibinfo{author}{\bibfnamefont{D.}~\bibnamefont{Joubert}},
\bibinfo{title}{\bibfnamefont{From ultrasoft pseudopotentials to the projector augmented-wave method}},
\bibinfo{journal}{Phys. Rev. B} \bibinfo{volume}{59}
(\bibinfo{year}{1999}) \bibinfo{pages}{1758-1775}.

\bibitem{pw91}
\bibinfo{author}{\bibfnamefont{J.P.} \bibnamefont{Perdew}},
\bibinfo{author}{\bibfnamefont{Y.}~\bibnamefont{Wang}},
\bibinfo{title}{\bibfnamefont{Accurate and simple analytic representation of the electron-gas correlation energy}},
\bibinfo{journal}{Phys. Rev. B} \bibinfo{volume}{45}
(\bibinfo{year}{1992}) \bibinfo{pages}{13244-13249}.

\bibitem{jiang03}
\bibinfo{author}{\bibfnamefont{D.E.} \bibnamefont{Jiang}},
\bibinfo{author}{\bibfnamefont{E.A.}~\bibnamefont{Carter}},
\bibinfo{title}{\bibfnamefont{Adsorption and diffusion energetics of hydrogen atoms on Fe(110) from first principles}},
\bibinfo{journal}{Surf. Sci.} \bibinfo{volume}{547}
(\bibinfo{year}{2003}) \bibinfo{pages}{85-98}.

\bibitem{monkhorst}
\bibinfo{author}{\bibfnamefont{H.J.} \bibnamefont{Monkhorst}},
\bibinfo{author}{\bibfnamefont{J.D.}~\bibnamefont{Pack}},
\bibinfo{title}{\bibfnamefont{Special points for Brillouin-zone integrations}},
\bibinfo{journal}{Phys. Rev. B} \bibinfo{volume}{13}
(\bibinfo{year}{1976}) \bibinfo{pages}{5188-5192}.

\bibitem{ferriani10tip}
\bibinfo{author}{\bibfnamefont{P.} \bibnamefont{Ferriani}},
\bibinfo{author}{\bibfnamefont{C.} \bibnamefont{Lazo}},
\bibinfo{author}{\bibfnamefont{S.} \bibnamefont{Heinze}},
\bibinfo{title}{\bibfnamefont{Origin of the spin polarization of magnetic scanning tunneling microscopy tips}},
\bibinfo{journal}{Phys. Rev. B} \bibinfo{volume}{82}
(\bibinfo{year}{2010}) \bibinfo{pages}{054411}.

\end{thebibliography}
\end{document}